\documentclass[aps,twocolumn,superscriptaddress,showpacs,floatfix]{revtex4}
\usepackage{epsfig}
\usepackage{graphicx}
\usepackage{bm}
\usepackage{natbib}
\usepackage{dcolumn}
\usepackage{amsmath}
\usepackage[english]{babel}
\usepackage{flafter}
\usepackage{wrapfig}

\usepackage{float}
\newfloat{figtable}{t}{lop}
\floatname{figtable}{Table}

\begin{document}

\title{Diffusion of scientific credits and the ranking of scientists}

\author{Filippo Radicchi}
\affiliation{Complex Networks and Systems, Institute for Scientific Interchange (ISI), Torino, Italy}

\author{Santo Fortunato}
\affiliation{Complex Networks and Systems, Institute for Scientific Interchange (ISI), Torino, Italy}

\author{Benjamin Markines}
\affiliation{Center for Complex Networks and Systems Research (CNetS), School of Informatics and Computing,
Indiana University, USA}

\author{Alessandro Vespignani}
\affiliation{Center for Complex Networks and Systems Research (CNetS), School of Informatics and Computing,
Indiana University, USA}
\affiliation{Complex Networks and Systems, Institute for Scientific Interchange (ISI), Torino, Italy}

 %

\begin{abstract}
Recently, the abundance of digital data enabled the implementation of graph based ranking 
algorithms that provide system level analysis for ranking publications and authors.  
Here we take advantage of the entire Physical Review publication archive ($1893$-$2006$) to 
construct authors' networks where weighted edges, as measured from opportunely normalized citation counts, 
define a proxy for the mechanism of scientific credit transfer. On this network we define a ranking method 
based on a diffusion algorithm that mimics the spreading of scientific credits on the network.
We compare the results obtained with our algorithm with those obtained by local measures 
such as the citation count and provide a statistical analysis of the assignment of major career 
awards in the area of Physics. A web site where the algorithm is made available to perform 
customized rank analysis can be found at the address {\tt http://www.physauthorsrank.org}.
\end{abstract}

\maketitle


\section{Introduction}

Recently, the recording of social interactions and data in the electronic format has made 
available datasets of unprecedented size. This is particularly evident for bibliographic 
data whose study has received a boost from the information technology revolution and the 
digitalization process.
This has led to the definition of ranking measures which are supposed to provide objective and 
quantitative  measures of the importance of journals, papers, programs, people and disciplines~\cite{egghe,garfield}. While the validity 
of these metrics is object of debate~\cite{adler08}, it is now standard practice to  
consider measures such as the impact factor, the number of citations
and the h-index~\cite{hirsch05} to assess 
the scientific research production of individuals and institutions. 
In this context the use of multipartite networks 
as the natural abstract mathematical representation of the data is particularly convenient 
and several studies have recently focused on the study of co-authorship networks, paper citation 
networks, etc.~\cite{Newman:2001,Newman:2001a,Newman:2001b,vicsek:02,Redner98}. In general, each of these networks 
is an appropriate bipartite or unipartite network projection of the original bibliographic dataset 
where authors and papers are nodes and citations, authorship and other bibliographic information 
define the links among nodes~\cite{Redner98, chen07}.

The possibility of a system level study of these networks has opened new possibilities 
for the bibliometric analysis aimed at evaluating the impact of scientific collections, 
publications and scholar authors. In particular, the field has leveraged on graph based 
ranking algorithms developed in the context of the 
World Wide Web~\cite{brin98,kleinberg99,donato,sidiropoulos,walker07} to 
provide the impact and prestige of papers and authors. The final goal
of ranking bibliographic data is even more ambitious as it ultimately
concerns the possibility of predicting the
evolution of impact and ranks on the basis of past data~\cite{donato}.

Criticisms to the ranking mechanism are generally rooted in the fact
that the common indicators, like the simple citation counts or the
metrics derived from this quantity, do not truly account for the
actual merit of a scientist. Citations have different values
depending on who is the citing scientist, defining a complicated
mechanism of scientific credit diffusion from author to author. Even at 
the simplest level, this is a very non-local process in which
scientists endorse each other through the
process of citing each other's works. In order to take
into account this perspective, we have defined an
approach that bases the author's ranking on a diffusion
algorithm that mimics the diffusion of scientific credits along time.
Here we take advantage of the set of all $407\,236$ papers published between $1893$ and $2006$ 
in journals of the Physical Review (PR) collection (see section~\ref{sec:dataset} for a detailed 
description of the set). This collection is surely an exceptional proxy of the activity 
in the physical sciences and the impact that individual scientists have generated in the 
field~\cite{redner05}. The PR dataset has been already exploited to analyze paper citation network and measure 
the impact of a specific paper both with local (individual
paper/author) metrics (number of citations) and with graph-based 
ranking algorithms~\cite{chen07,walker07}. 
Here we propose a system level algorithm with the aim of ranking
authors by mimicking the scientific credit spreading process.
We first construct an author-to-author 
citation network that fully accounts for the bibliometric data 
relative to the credit given from any author to other authors. We then
define an appropriate graph-based ranking algorithm that simulates the
diffusion of  credits exchanged by the authors over the
whole network. The algorithm takes into account that
citations from more important authors have higher relevance than
citations from less important authors and the non-local nature of the
diffusion process in which any author can in principle 
impact the score of far away nodes through the diffusion process.
Finally, the proposed ranking technique is compared with other commonly used methods, 
which are based only on local properties of the citation network.

The paper is organized as follows. We first give a brief description of the PR dataset 
(section~\ref{sec:dataset}).  In section~\ref{sec:netw} the weighted
citation network between authors
is defined and analyzed. The description of the Science Author Rank Algorithm (SARA) is performed in section~\ref{sec:rra}. This algorithm is used 
for the estimation of the scientific impact of physicists along time. We compare SARA 
with other ranking schemes like Citation Count and Balanced Citation Count 
in section~\ref{sec:compare}. In section~\ref{sec:prediction},  we test SARA by using the list of the  
winners of the major prizes in Physics. This list of prominent physicists is in fact 
the best benchmark on which we may test our algorithm. We finally conclude and report final comments in 
section~\ref{sec:conclusions}.

\section{Description of the dataset}
\label{sec:dataset}

Our database is composed of the set of all $407\,236$ papers published between $1893$ and $2006$ in journals 
of the collection of Physical Review (PR). 
The journals considered here are Physical Review Series I, 
Physical Review, Physical Review A, Physical Review B,  Physical Review C,  Physical 
Review D,  Physical Review E,  Physical Review Letters  and Reviews of Modern Physics. 
For each paper the editorial office of PR provided an {\it xml} file from which we can extract 
the names of its author(s), date, journal, volume and page of publication, its references, 
the PACS~\footnote{PACS stands for Physics and Astronomy Classification Scheme. This scheme is nowadays 
universally adopted by the majority of Physics journals in order to well classify papers. Since $1980$, 
Physical Review's journals have started to associate a set of PACS
numbers (on average three PACS numbers per paper) with every published
paper.} numbers and other additional information.

The list of references at the end of each paper allows to construct a
network of citations between papers. According to our database, the
total number of references (obtained by summing all references over
all papers) is $9\,359\,556$ of which $3\,866\,471$~
\footnote{Actually, the total number of internal references reported
  by the PR database is $3\,866\,822$, but $351$ of them are clearly
  wrong since they refer to  papers citing newer papers (i.e., the
  year of publication of the citing paper is smaller, in some case
  even of $30-40$ years, than the one of the cited paper). We cannot
  {\it a priori}  exclude the possibility of other wrong internal
  references, but there is no other simple method to determine whether
  a reference is good or not.}  are internal references (i.e.,
references to papers appeared in PR journals). 

In this work we have neglected all references of the type ``First
author {\it et al. }'' and all references pointing to papers written
by authors without any publication in the PR journals.  Using these
criteria, we identify $8\,783\,994$ total references (including the
$3\,866\,471$ internal references). 

In the rest of the paper and all our analysis, we consider all
$8\,783\,994$ references. As already stated, these references
include all papers, published or not in PR journals, referenced by
papers published only in PR journals.

\section{Construction of the weighted author citation network}
\label{sec:netw}

\begin{figure}[!htbp]
\includegraphics[width=0.45\textwidth]{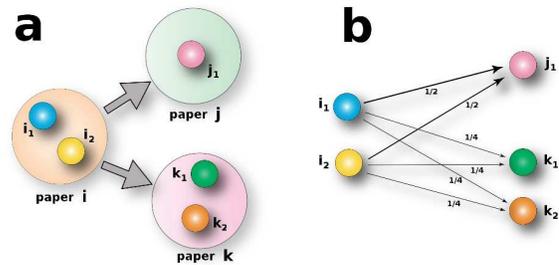}
\caption{(Color online) Projection of the PCN into a WACN. (a) In the network of
  citations between papers, the article $i$, written by two authors
  $i_1$ and $i_2$, cites two papers $j$ and $k$, written
  by one author $j_1$ and two co-authors $k_1$ and $k_2$, respectively. (b) The WACN
  is then simply generated by connecting  with a directed link both
  $i_1$ and $i_2$ to $j_1$, each with weight $1/2$, and to $k_1$ and
  $k_2$, each with weight $1/4$.}
\label{fig1_new}
\end{figure}

A weighted
citation network between authors (WACN) can be easily determined as a particular projection of the paper citation network (PCN)
constructed by the list of references described in
section~\ref{sec:dataset} [see Figure~\ref{fig1_new}]. 
Consider for
instance a paper $i$, written by the $n$ co-authors 
$i_1$, $i_2$, \ldots , $i_n$, which cites a paper $j$, written by the
$m$ co-authors $j_1$, $j_2$, \ldots , $j_m$.  A natural way to project
the unweighted directed link $i \to j$  between papers $i$ and $j$
into a WACN is to create $n \cdot m$ directed  connections from each
of the $n$ citing authors to every of the $m$ cited authors (i.e., 
$i_k \to j_s \;, \forall k=1,\ldots,n$ and $\forall s=1,\ldots,m$), where every connection has weight 
equal to $w_{i_k,j_s}=1/\left( n \cdot m\right)$. Given a set of references 
(i.e., directed links between papers), the weight of a directed link
between two authors  will be the sum of all the weights over all the
references in the
set. 

It is important to stress here that while the list of references does
not have ambiguity, the analysis of the author projection opens the
issue of names disambiguation. Indeed, common names may refer to
different authors and not all authors report
their full names in publications. In other words we could have a
multiplicity of authors identified by the same identifier. In
appendix~\ref{sec:netw_id} we provide a detailed analysis of this
and other related problems which are common issues in bibliometry.

\begin{figure*}[!htbp]
\includegraphics[width=0.9\textwidth]{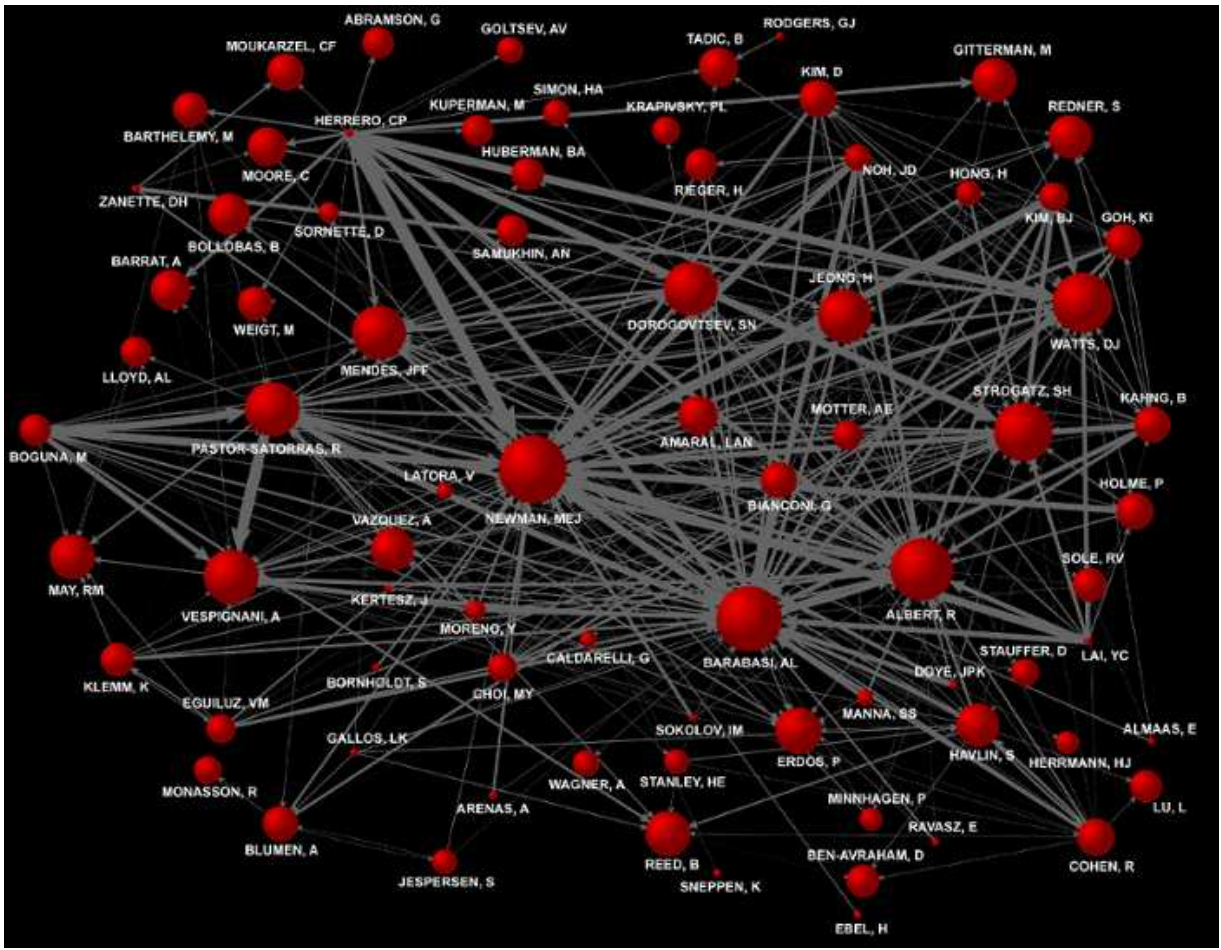}
\caption{(Color online) We generated the
citation network based on all papers published in PR journals about
the topic ``complex networks''. For clarity, only links with weight
above a certain threshold have been plotted. As a consequence only
top-physicists in this field are shown. The width of each connection
is proportional to its weight
and the size of the nodes is proportional to the sum of all weights of
incident links.}
\label{fig_network}
\end{figure*}

As an example of the network construction, in Figure~\ref{fig_network} we show the WACN of
the top-scientists in the field of ``complex networks''. In order to
construct this network, we first select out of the PR dataset only
papers whose titles contain keywords as ``complex network'',
``scale-free network'', ``small-world network'', etc. We then
consider their references and based on this list we project the PCN
into a WACN.

\subsection{Dynamical Representation of the Weighted Author Citation Network}
\label{sec:time}
In principle, a single WACN may be constructed based on the full set
of  the $8\,783\,994$ total references described in
section~\ref{sec:dataset}.  This is however not very informative as
very old citations are mixed with new ones,  discounting the dynamical
information contained in the longitudinal nature of the database. In
addition, the rate of citation per unit time is steadily increasing
along the years. For this reason,  we define dynamical slices of the
database containing the same number of citations. We first sort the
full list of references according to their date (i.e., the date of the
publication of the citing paper). Then we divide this list in $M_I$
homogeneous intervals, where  homogeneous stands for intervals with
the same number of references $M_R$.  In order to avoid abrupt
changes, we consider overlapping intervals, in the sense that the
$q$-th interval shares its first $M_R/2$ references with the
$(q-1)$-th interval and its last $M_R/2$ references with the
$(q+1)$-th interval.  It should be noticed that  this sharp division
may split references of the same citing paper into different
contiguous intervals, but this ``border effect'' may be considered
negligible since we  consider $M_R$ much larger than the average
number of references per paper (all results have been obtained by
using $M_I=39$ and $M_R=488\,000$, while on average each paper has
$20-30$ references). Moreover, we should remark that we  can relate
each interval with real time by simply associating the average of the
dates of all the references belonging to the interval with the
interval itself. However, since the rate of citation per unit of time
is increasing almost exponentially with time, the homogeneity of
references in each interval does not correspond to homogeneity in
time: for instance the first interval spans more than $70$ years
of publications ($1893$-$1966$), while the last interval is
representative for the publications of only one year ($2006$). The
choice  $M_R=488\,000$ adopted in this paper ensures that intervals
are representative of periods of time not shorter than one year.

\subsection{Properties of the Weighted Author Citation Network}
\label{sec:netw_prop}

We provide in this section a simple statistical analysis of the WACNs. 
In particular we monitor the number of authors and their indegree 
and instrength distributions, where for example the instrength of a node $i$ 
is defined as 
\begin{equation}
s_{i}^{in} = \sum_j w_{ji} \;\;\;,
\label{eq:instr} 
\end{equation}
i.e., the sum of all weights of the links pointing to $i$~\cite{barrat04}.
First of all, it is interesting to note that quantitatively the
properties of the WACNs are not constant in time. This is understandable
since the production  of scientists has strongly changed
during the last century.

\begin{figure}[!hbt]
\includegraphics[width=0.45\textwidth]{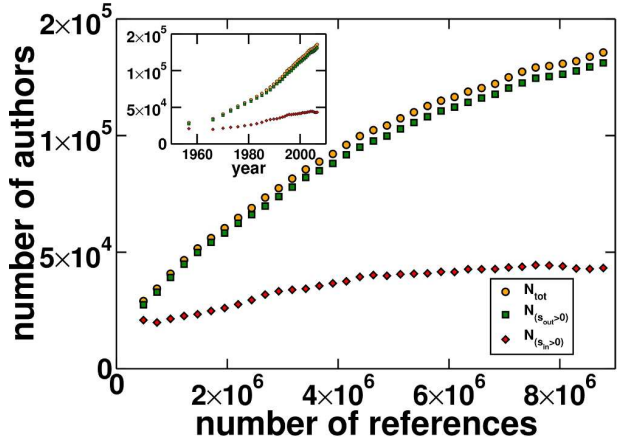}
\caption{(Color online) In the main plot, the total number of authors $N_{tot}$ (yellow circles),
  number of authors with outstrength larger than zero
  $N_{(s^{out}>0)} =\sum_j \theta \left(s_{j}^{out} \right)$ (green squares) and number of authors with instrength larger than
  zero $N_{(s^{in}>0)}=\sum_j \theta \left(s_{j}^{in} \right)$ (red diamonds) are plotted as  functions of the number of
  references (referenced papers), where $\theta\left(\cdot\right)$ is the step function equal to one when its argument is 
larger than zero and null otherwise. In the inset the same quantities as those of the main
  plot are considered, but now they are plotted as functions of time. More specifically,
  each $x$-value corresponds to the average publication year of papers belonging to the respective
  dynamical slice of the main plot.}
\label{fig2_new}
\end{figure}

\begin{figure}
\includegraphics[width=0.45\textwidth]{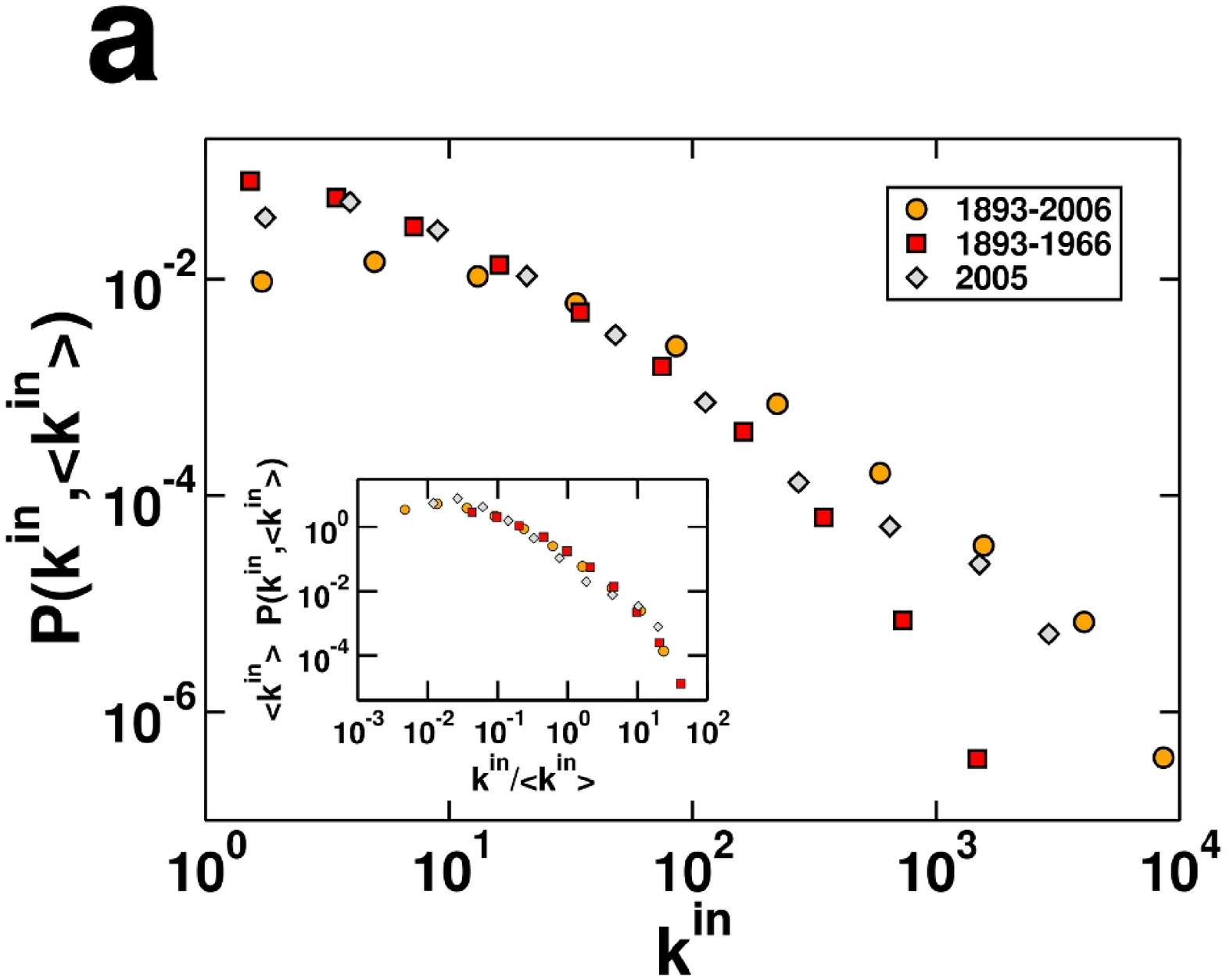}
\\
\includegraphics[width=0.45\textwidth]{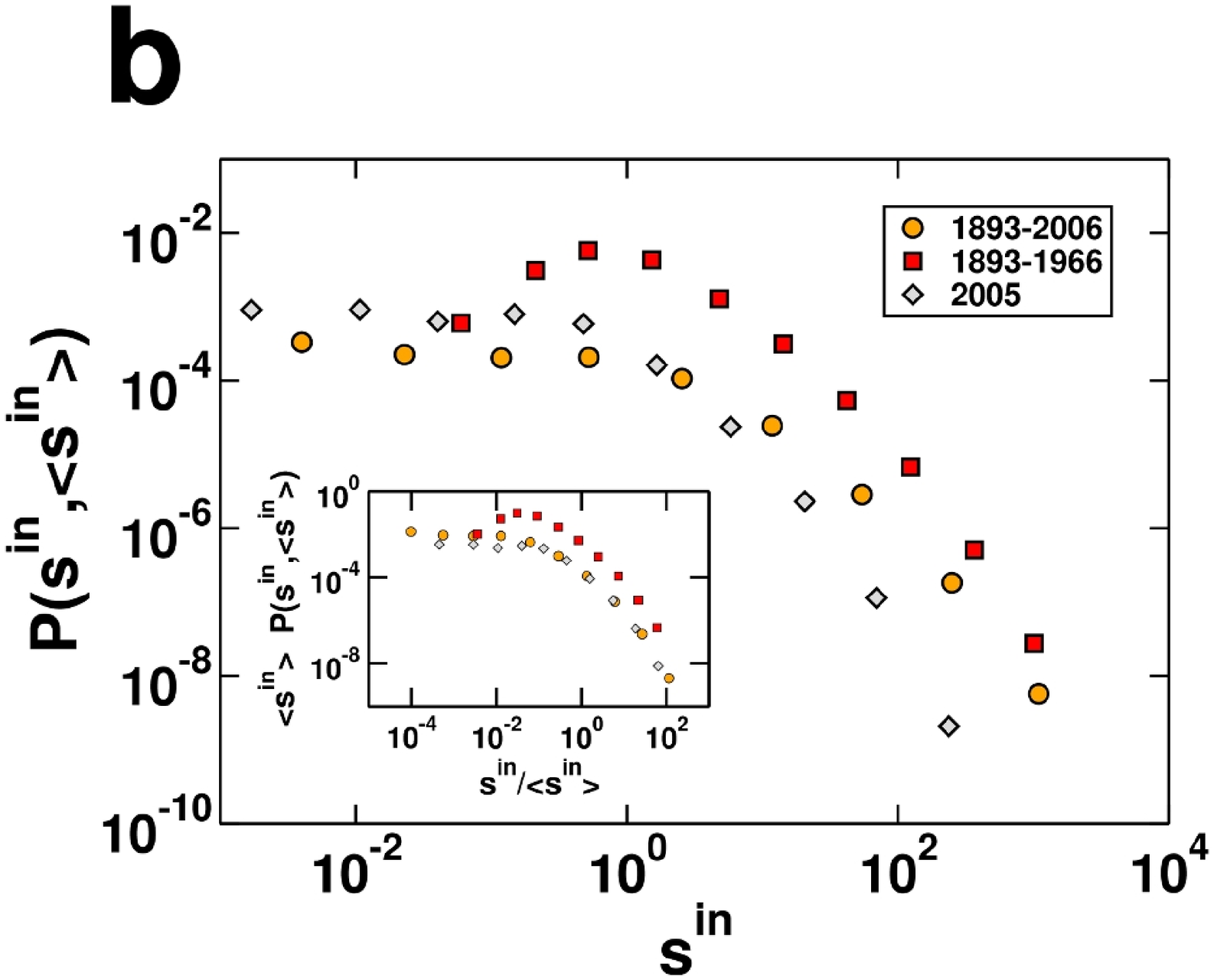}

\caption{(Color online) Probability densities for the indegree (a) and the
  instrength (b). Calculations have been performed on different WACNs
  based on papers published in different periods of time (yellow circles $1893-2006$, red squares $1893-1966$, gray diamonds $2005$). The insets
  show the same distribution as in the main plots, but opportunely
  rescaled by their average values.}
\label{fig3_new}
\end{figure}

From Figure~\ref{fig2_new}, one can
qualitatively appreciate the former observation: the total number of
nodes in the network (i.e., the number of scientists citing or cited
in a particular period of time) is an increasing function of time. It
should be stressed that this behavior is mainly a consequence of
the increment of scientists in physics as one can deduce from the
time-increment of the number of nodes with non-zero instrength (i.e.,
cited authors) that is growing in a much slower fashion.
\\
The indegree
distributions calculated on different WACNs are generally
different. Nevertheless, if we consider the relative indicator given
by the ratio of the  citing authors ($k^{in}$) to a scientist in
a given WACN divided by the average number ($\langle k^{in} \rangle$)
of citing authors over all  physicists in the same WACN, the distributions
of the rescaled variable $k^{in}/\langle k^{in} \rangle$ obey the same
universal curve [see Figure~\ref{fig3_new}a]. This result is in accordance with  the remarkable
scaling recently discovered on PCNs~\cite{radicchi08}. The same is not valid for
the instrength distribution since a simple scale transformation does
not seem to lead to a universal behavior.

\section{Science Author Rank Algorithm}
\label{sec:rra}

The author-to-author network can be used to define a graph based ranking algorithm that uses the 
global features of the network to account for the impact of each author. Analogously to 
various ranking algorithms such as PageRank~\cite{brin98}, CiteRank~\cite{walker07}, the HITS scores~\cite{kleinberg99}, 
etc., we define an iterative algorithm 
based on the notion of diffusing scientific credits. In practice we imagine that each author 
owns a unit of credit which is distributed to its neighbors proportionally to the weight of 
the directed connection. Each author thus receives a credit that is then redistributed 
to neighbors at the next iteration and so on. In other words, the SARA
simulates the diffusion of credits on the global network
according 
to a diffusion probability proportional to the weight of the links. 
\\
Let us be more specific. Once the WACN has been defined as detailed in
section~\ref{sec:netw}, we calculate the SARA score for each node $i$
according to

\begin{equation}
P_i = \left(1-q\right)\sum_j \; \frac{P_j}{s^{out}_j} w_{ji} +  q
z_i+ \left(1-q\right) z_i \sum_j \; P_j \; \delta \left( s^{out}_j
\right).
\label{eq:pg}
\end{equation}
Here $P_i$ is the score of the node $i$, $1 \geq q \geq 0$
is the damping factor, $w_{ji}$ is the weight of the directed
connection from $j$ to 
$i$, $s^{out}_j$ is the outstrength of the node $j$ (i.e., the sum of the weights of 
all the links outgoing from the $j$-th vertex, $s^{out}_j=\sum_k
w_{jk}$) and finally $\delta (x)=1 ,$ if $x=0$ and $\delta (x)=0 ,$
otherwise. The first term on the r.h.s. of Eq.(\ref{eq:pg}) represents
the diffusion of credit through the network: scientist $i$
receives a portion  of credit from each citing author $j$ and each
amount of credit is linearly proportional to the weight $w_{ji}$ of the arc
linking $j$ to $i$. The second and the third  terms stand
from the redistribution of credits to all scientists in the
network. A portion $q$ of the credit of each node is redistributed 
to everyone else (i.e., second term), with the exception of 
dandling ends (i.e., nodes with null outstrength), which distribute their whole credit 
(i.e., third term). The meaning of the
redistribution of credit is that everyone is in ``scientific debit''
with the whole scientific community, since a general background is at
the basis of the knowledge of every scientist.  In particular, the
credit is distributed homogeneously among papers in the network. The
factor $z_i$ takes into account the normalized scientific credit given
to the author $i$ based on his productivity. $z_i$ is calculated
according to the formula

\begin{equation}
z_i = \frac{\sum_p \delta_{p,i} \; 1/n_p}{\sum_j \sum_p \delta_{p,j} \; 1/n_p} \;\;\; ,
\label{eq:credit}
\end{equation}
where $p$ represents the generic paper $p$ and $n_p$ the number of
authors who have written the paper $p$. Moreover,
$\delta_{p,i}=1$ only if the $i$-th author wrote the paper $p$, otherwise it equals zero. The sum runs over 
all different papers (citing and cited). Basically, each paper
receiving a credit is going to redistribute it equally among all
co-authors of the paper. The fact that the $z_i$s are not homogeneous
(differently from the original formulation of PageRank~\cite{brin98},
where $z_i=1/N \;,\; \forall \; i$ with $N$ total number of authors)
is of fundamental importance: each paper is carrying the same amount
of knowledge independently of the number of co-authors. The
denominator of the r.h.s. of Eq.(\ref{eq:credit}) serves only for
normalization purposes. 
\\ 
The stationary values of the $P_i$s can be easily computed
recursively, by setting at the beginning $P_i=z_i \; , \; \forall i$ 
(but the results are independent of the choice of the initial values) and iterating
Eqs.(\ref{eq:pg}) until they converge to values stable 
within {\it a priori} fixed precision \footnote{If $t$ stands for the stage of convergence, this means 
$\left| P^{(t-1)}_i -  P^{(t)}_i \right| < \epsilon \; , \; \forall \; i$, where $\epsilon$ represents 
the {\it a priori} fixed precision. Here we set $\epsilon = 10^{-6}$; typically $20-30$ iterations 
are needed for convergence.}.
\\
The scores calculated according to Eq.(\ref{eq:pg}) depend on the
particular value chosen for the damping factor $q$. In all results
shown in this paper, we always set $q=0.1$. This is the value for
which the predictive power of SARA is maximized. An exploration of the
dependence of the predictivity of SARA as a function of the damping
factor $q$ is reported in Appendix~\ref{appendix}.

\subsection{Ranking Authors}

\begin{figure}[!hb]
\includegraphics[width=0.45\textwidth]{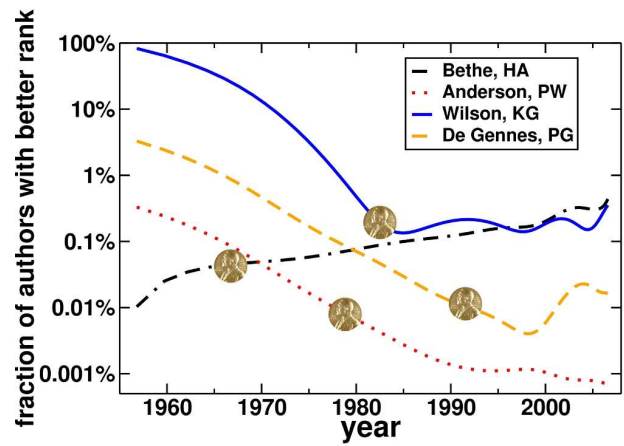}
\caption{(Color online) Evolution of the relative rank expressed as
top percentile  of four Nobel
  laureates: ``Bethe, HA'' ($1967$, black solid line), ``Anderson, PW'' ($1977$, red dotted line),
  ``Wilson, KG'' ($1982$, blue solid line) and ``De Gennes, PG'' ($1992$, yellow dashed line). Scientific
  merit is quantified by using Eq.(\ref{eq:prob}), which counts the
  author's percentile as the relative number of authors with better rank
  than the considered scientist. The figure shows how
  relative rank is related in time with the Nobel prize (date
  of the award indicated by the symbol). The diagram monitors the
  scientific carrier of the awardees, essentially from the beginning,
  with the only exception of ``Bethe, HA'', whose activity began much
  earlier than that of the other three scientists.}
\label{fig5_new}
\end{figure}

\begin{figure*}[!ht]
\vskip .7cm
\includegraphics[width=0.45\textwidth]{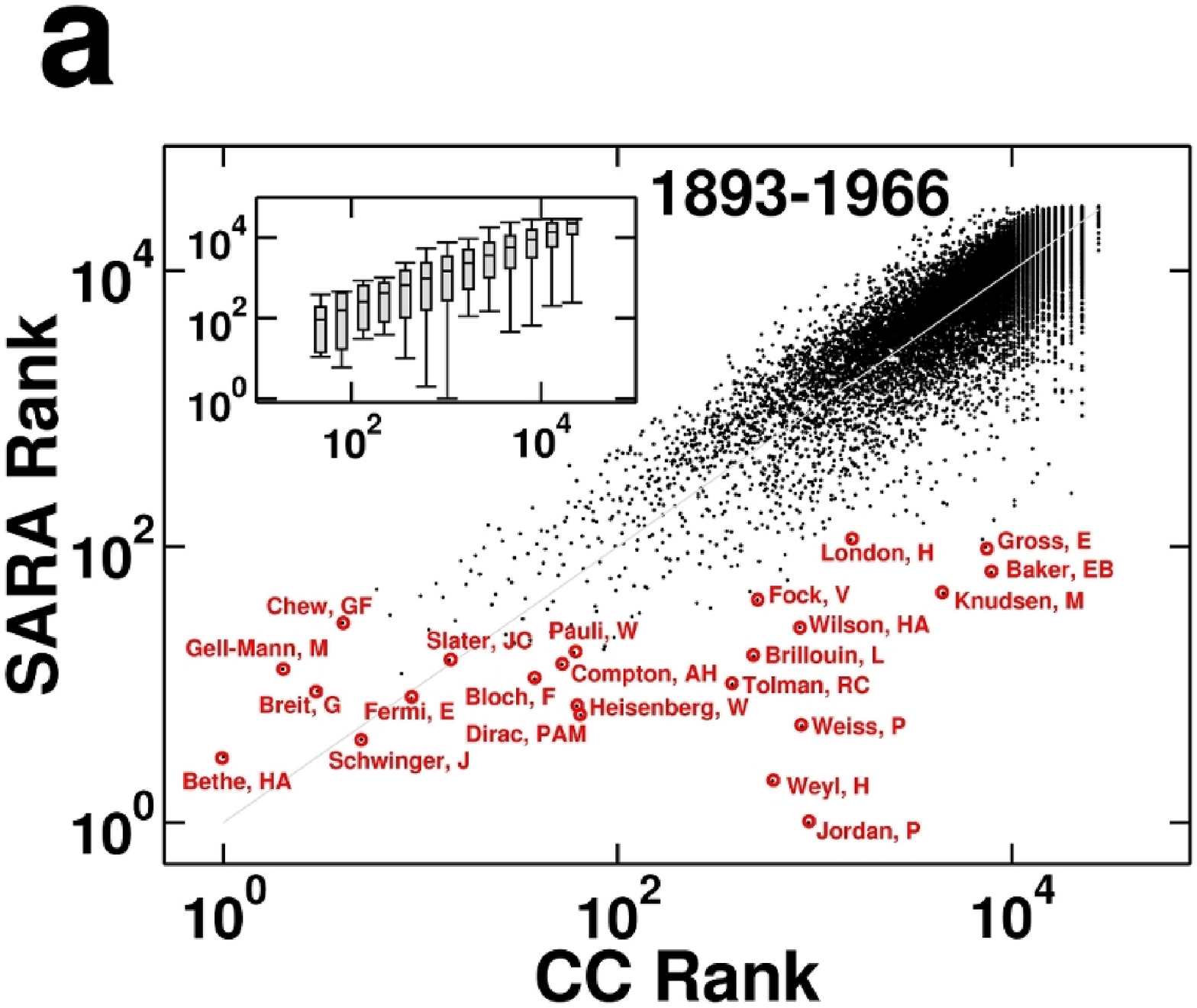}
\quad
\includegraphics[width=0.45\textwidth]{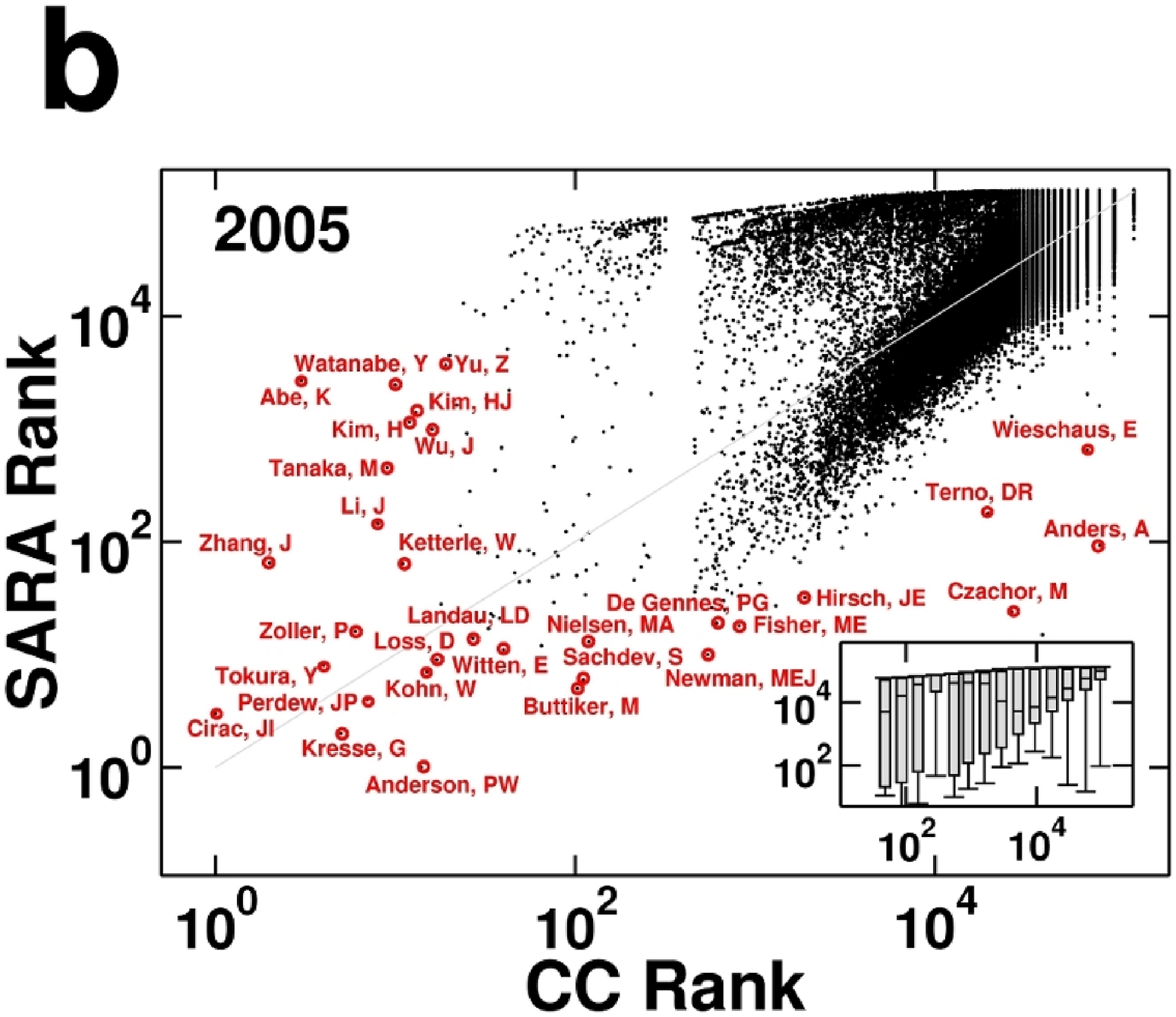}
\vskip .1cm
\includegraphics[width=0.45\textwidth]{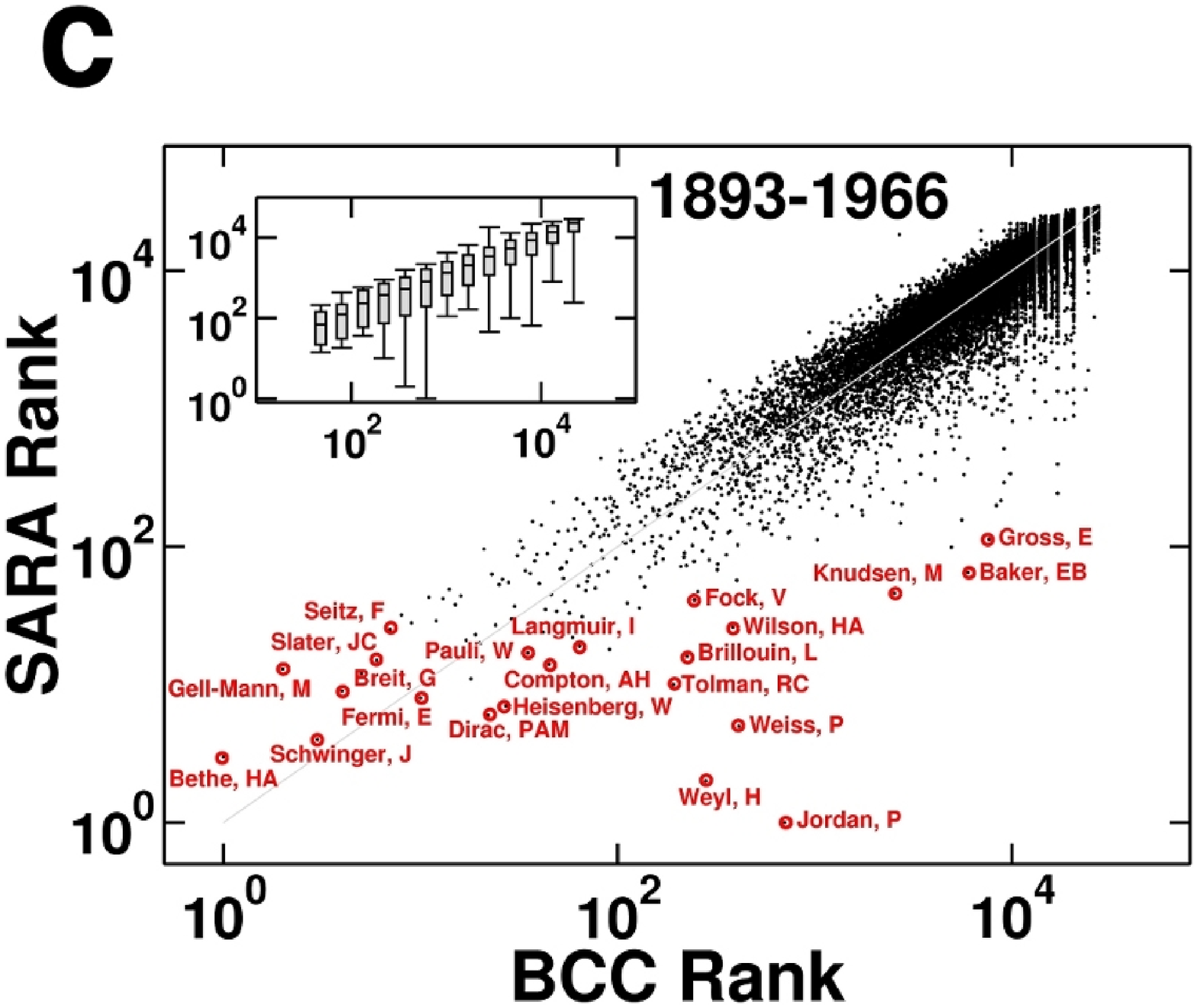}
\quad
\includegraphics[width=0.45\textwidth]{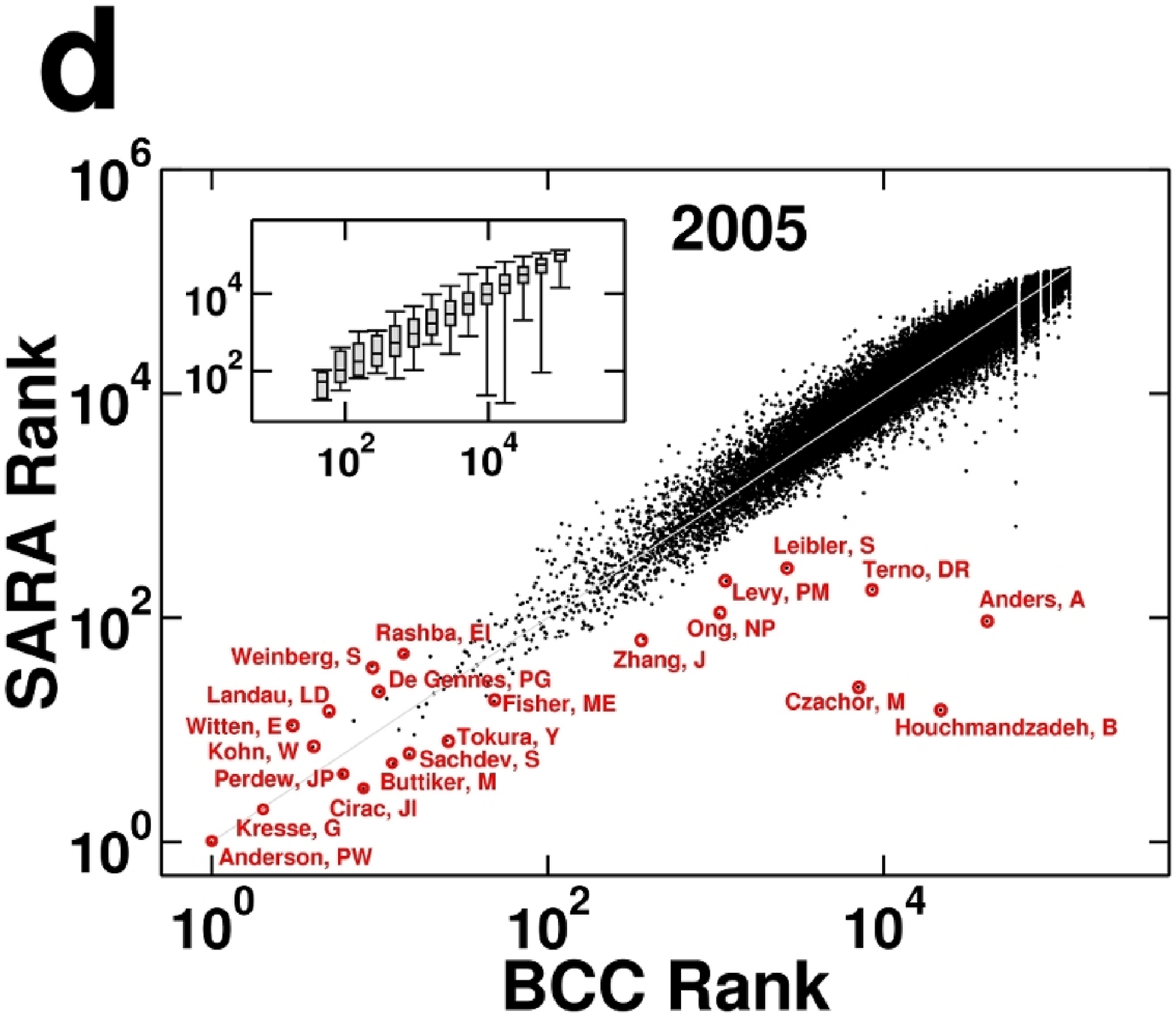}
\caption{(Color online) Scatter plots of SARA rank {\it versus} CC rank
    [(a) and (b)] and BCC rank [(c) and (d)]. 
Plots in (a) and (c) refer to the author citation network based on papers published between $1893$ 
and $1966$, while plots in (b) and (d) have been generated by using the author citation network based 
on papers published in $2005$. In all insets, the same data as the ones analyzed in the respective 
main plots have been logarithmically binned. For each bin we plot maximum and minimum values 
(error bars), $90\%$ confidence intervals (boxes) and median
(horizontal bars inside boxes) of the SARA rank. In all plots, outlier
points stress the most significant differences between SARA and the
other techniques. Authors badly ranked in CC or BCC methods and
well classified in SARA are generally very prominent physicists. By looking at
figures (a) and (c) for example, we see scientists of the caliber of
``Jordan, P'' and ``Weyl, H'' occupy the top-positions in SARA ranking,
while their ranks are two orders of magnitude smaller according
to  CC or BCC methods. On the other hand, the majority of authors
poorly ranked by the SARA technique and well ranked by CC method correspond
to poorly defined identifiers referring in general to multiple physical persons
[see figure (b)]: names like ``Li, J'' or ``Yu, Z'' are very common in
China and for this reason their CC score is very high; SARA differently
is able to capture the low scientific relevance of all these authors,
ranking them at positions about three orders of magnitude higher than
the ones obtained with the CC method.}
\label{fig6_new}
\end{figure*}

The SARA is used to provide a ranking of the authors in the PR
database. Given an author-to-author network, we calculate the score
of each author according to Eq.(\ref{eq:pg}) and assign a rank
position to this scientist. The higher is the score of a scientist,
the higher is her/his rank. As described in section~\ref{sec:netw}, we
decided to preserve the longitudinal nature of the PR database and
construct WACNs corresponding to dynamical slices of the database
containing the same number of citations. In this way we 
can have a dynamical perspective on the evolution of the merit of authors along the years.
\\
As prototypical examples, we show in Figure~\ref{fig5_new} the
evolution of the relative rank of four Nobel Laureates. For each author $i$
we calculate its relative rank as 
\begin{equation}
R_i = 1/N \; \sum_{j \neq i} \theta \left( P_j - P_i \right) \;\;\; ,
\label{eq:prob}
\end{equation}
which basically stands as the probability to find an author with better score than  author 
$i$. $N$ is the total number of authors in the WACN, while the step
function $\theta(\cdot)$ is equal to one only when 
its argument is equal to or larger than one, otherwise it is zero. 
The relative rank in other words defines the
top percentile of each scientist. It should be
stressed that the relative rank of Eq.(\ref{eq:prob}) works better
than the absolute one in the case of comparison of scientific
performances in different historical periods, since the number of
authors in the WACN is increasing rapidly 
in time (see Figure~\ref{fig2_new}).

From  Figure~\ref{fig5_new}, we can clearly see that relative rank
 dynamics of Nobel laureates is qualitatively related in time with the
 achievement of the prize:
top-performances are reached close to the date of the assignment of
the honor. Indeed, it is worth remarking that the method naturally 
accounts for the fact that the rate of
 citations per unit time is steadily increasing through the years by
 defining dynamical slices of the database containing the same number
 of citations. Discounting old citations, the author's rank
 becomes a dynamical quantity that changes according to the author's
 research activity as well as the success of new research
 fronts. Thus, rank is related to the actual impact of the
 research of an author at a given time and is changing through the
 years.

\section{Comparison with Different Metrics}
\label{sec:compare}

Assessing the reliability and the results of any ranking method is not easy. The main 
question is to which extent the SARA algorithm is providing a better rank than other ranking 
methods commonly used in scientific impact analysis. For this reason, we consider two basic 
measures which are commonly used to rank authors. The first is the Citation Count (CC) with 
which authors are simply ranked by the total number of citations
received in a given time window (note that the number of citations
does not correspond to the indegree of the author in the citation
network). CC is traditionally the simplest and mostly used quantity
for measuring the scientific impact: popular indicators, as the
h-index~\cite{hirsch05} for instance, are based on this simple
metrics.
The second measure is the Balanced Citation Count (BCC) that discounts the effect of multiple 
authored papers in the citation count by normalizing the citation weight by the total number of 
authors of the cited paper [i.e., authors are ranked on the basis of their instrength as defined in 
Eq.~(\ref{eq:instr})].  As a first comparison of the rankings obtained with the three different 
methods, we show in Figure~\ref{fig6_new} the scatter plot in which each author is identified by its 
SARA ranking and CC or BCC rank. If the methods provide the same ranking all the points would 
fall on the diagonal. Fluctuations are indicated by the cloud of the scattered plot about 
the line indicating the linear behavior. Indeed, it is possible to show that,
in the absence of degree-degree correlations in the network, diffusion algorithms such as the SARA 
are providing a score that is on average proportional to the indegree dependence of the diffusion 
process~\cite{fortunato08}.
However, important fluctuations appear: some nodes can 
have for example a low SARA rank despite a modest indegree, whereas some others can have 
a surprisingly large SARA despite a high indegree, as it is possible to see in Figure~\ref{fig6_new}.
We believe that the potential refinement offered by this method is its ability to uncover 
such outliers. It is interesting to see that most of the outliers corresponding to authors badly 
ranked with the CC and BCC methods are indeed very important scientists that are highly ranked with our method.

\section{Benchmarking the Science Author Rank Algorithm}
\label{sec:prediction}

\begin{figure}[!htb]
\vskip .7cm
\includegraphics[width=0.45\textwidth]{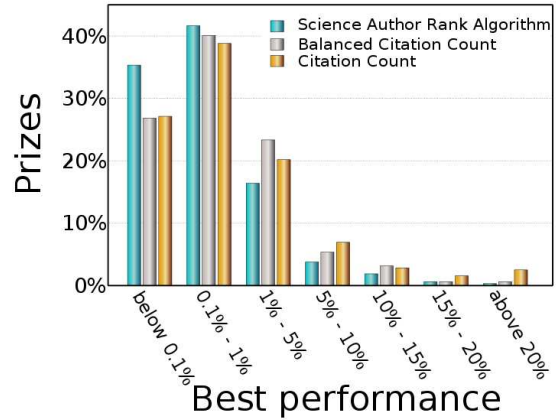}

\caption{(Color online) We consider some of the main prizes in Physics (Nobel prize, Wolf prize, Boltzmann medal, Dirac 
medal and Planck medal). To each prize, we associate the best performance of the scientist who earned 
that honor. The performance of an author at a given time is quantified
by the author's percentile defined as the percentage of other 
authors who have  a better rank at the same time [see Eq.~\ref{eq:prob}]: the lower is this percentage, the better is the 
performance of the considered scientist. SARA is more predictive than both CC and BCC: according to 
SARA ranking, the $35\%$ of the prizes have been assigned to scientists who have reached a position
below the $0.1\%$. The SARA tells that $77\%$ of the considered honors have been earned by scientists 
with a best performance rank lower than $1\%$. As term of comparison, 
according to CC (BCC) ranking the former rate decreases to $66\%$ and
$67\%$, respectively.}
\label{fig7_new}
\end{figure}

\begin{figtable*}
\vskip .7cm
\includegraphics[width=0.9\textwidth]{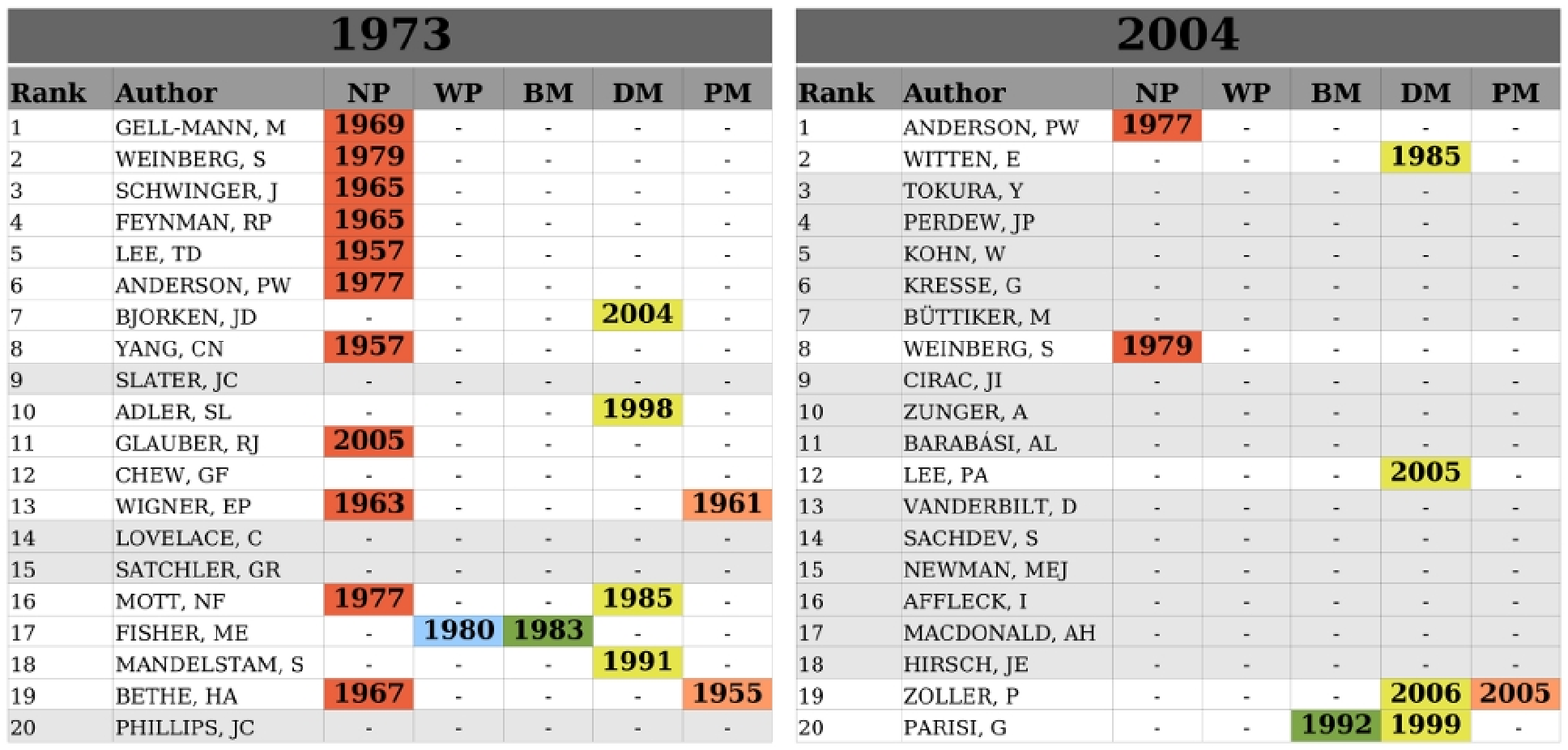}
\vskip .1cm
\caption{(Color online) Top $20$ scientists according to the SARA method. The rankings are determined 
by considering all papers published in the periods $1967$-$1973$ (left) and  
$2003$-$2004$ (right). We highlighted in gray scientists, who have not yet 
earned any of the major prizes [NP=Nobel Prize, WP=Wolf Prize, BM=Boltzmann Medal, DM=Dirac Medal, PM=Planck Medal]. "Kohn, W" has earned the NP in Chemistry in $1998$.}
\label{table}
\end{figtable*}

The previous analysis is not an accurate author by author analysis but a procedure to identify the 
most evident outliers. In order to produce a more refined analysis on the effectiveness of the SARA 
ranking, we test the predictive power of the three ranking methods by studying the assignment of major prizes 
and awards (in Ref.~\cite{garfield86} it has been already shown that scientists with high 
CC scores have high probability to earn a Nobel prize in their discipline). We expect that a 
better performing ranking would identify most of the award winning 
authors by placing those at very top ranks. In other words we assume that awards and prizes are an 
outcome of a peer performed rank analysis that singles out the most highly ranked authors. 
This human ranking process, obtained with the hard work of committees and the help (in many cases) of the whole community  
 can be considered as a benchmark for the ranking algorithms. We expect that the better 
the algorithm is performing, the more awarded authors will be found in the top rank brackets. 
In Figure~\ref{fig7_new}, we see how SARA improves the prediction in the assignments of 
 major prizes in Physics with respect to both CC and BCC methods. The probability to earn 
a prize is consistently higher for authors who have reached top rank 
positions~\footnote{The best performance $R^m_i$ of scientist $i$ is calculated according 
to $R^m_i = \min_t R_i\left(t\right)$, where $R_i\left(t\right)$ is the relative rank defined 
in Eq.(\ref{eq:prob}) of the $i$-th author in the WACN corresponding to the $t$-th time slice 
of the PR database.} according to SARA than for scientists who have occupied the same positions in CC or BCC rankings.

\

Finally, we provide a table [see Table~\ref{table}]
with best ranked scientists at the end of years $1973$ (period $1967$-$1973$) and $2004$ (period $2003$-$2004$), where  we single out 
those who have not yet received any of the major awards we considered in the present analysis. It is 
important to stress that some prizes are disciplinary and cannot apply
to all authors. Nevertheless, the majority of  the 
scientists ($16$ out of $20$) listed in the left part of table~\ref{table} (period $1967$-$1973$) have earned one of the prizes considered in this analysis. On the other hand, all scientists listed in the right part of table~\ref{table} (year $2004$) are, by our knowledge,
top-physicists in their field of research and probably eligible to
very important prizes in Physics not only in accordance with our
criteria.

\section{Conclusions}
\label{sec:conclusions}
In this paper we propose a new measure for ranking scientists
mimicking the spread of scientific credits among authors. 
The proposed technique, called Science Author Rank Algorithm (SARA), is
similar in spirit to the standard ranking procedure implemented
for pages in the 
World Wide Web~\cite{brin98}. SARA is based on a mixed process, where a biased random walk is combined with
a random distribution of the credits among the nodes.
On a global level, the algorithm takes into account that inlinks from
highly ranked authors are more important than inlinks from 
 authors with low rank and measures the non-local effects of the spreading of scientific credits 
into the network. The non-local characteristics of this algorithm are evident as any author 
can in principle impact the  score of far away nodes through the diffusion process 
and the fact that the  score of an author is more affected by the  score 
of its neighbors than the raw number of inlinks. 
\\
We apply SARA on Weighted Author Citation Networks (WACNs) directly
constructed from the paper citation network based on articles
published in the Physical Review (PR) collection between $1893$ and
$2006$. This large dataset allows the estimation through SARA scores of
the scientific relevance of physicists along time. The time behavior
can be monitored by simply using the longitudinal nature of the PR
database and therefore constructing WACNs representative of different
periods of time. A quantitative comparison between rankings obtained
via SARA scores or other more popular heuristics shows the great
improvement that can be obtained  by considering the whole citation
network instead of only its local properties.
\\
As practical application of our ranking recipe, we have developed a Web
platform ({\tt http://www.physauthorsrank.org}) where the evolution of the
scientific relevance of all physicists, with at least a publication in
PR journals before $2006$, can be plotted. The Web site offers several 
additional features such as the evaluation of the authors' rank in their specific topical area. 

While we believe that the methodology exemplified by our approach 
entails more information than the simple citation counts or
the metrics derived from this quantity, including the h-index and its
related measures, we want to be the first to
spell out clearly the many caveats deriving by a non-critical approach
to similar ranking approaches. First of all it is worth remarking that
the present algorithm takes into account only the PR dataset. While
this may be appropriate to rank authors within the physics community,
it is clear that it does belittle the rank of authors who have got a
large impact in other areas or disciplines. This problem might be
mitigated by the inclusion of other databases or very extensive
citation repositories. The inclusion of larger repositories however
would amplify the disambiguation problem and this endeavour might not
be straightforward. For this reason we have added to our web platform
the user disambiguation process. The hope is that a collaborative
web2.0 approach may help in achieving progressively cleaner
datasets. A similar procedure has been recently 
proposed by Thomson Reuters with the
web site {\tt http://www.researcherid.com}~\cite{Enserink09}, where authors are asked
to link their {\it ResearcherID} to their own articles.
Another issue is the fact that our scientific credit spreading is
considering credits and citations just as a positive indicator of
impact. It is debated in the community how to consider the effect of
the so-called negative citations aimed at contradicting previous
results or conclusions. This is however a very subtle point as it is
almost impossible to say to which extent this kind of citations are
negative. In many cases even flaws or error may have the merit to open
new direction of research or the path to novel approaches. While we
prefer not to enter this discussion here it has to be kept in mind
that our method could be extended to define negative scientific credit. 
A final warning is concerning the general use and exploitation of
the global ranking approaches. It is clear that the obtained ranking
is just an indicator and cannot embrace the multifaceted nature and
the many processes at the origin of authors' reputation. The obtained
ranking has therefore to be considered as an extra element to be used
with grain of salt and especially in terms of ``order of magnitude''
more than in absolute value.

\begin{acknowledgments}
This work is partially supported by
the Lilly Endowment grant 2008 1639-000. to A.V. 
the grant of the European Community number 238597 ICTeCollective to S.F..
We acknowledge the American Physical Society
for providing the data about Physical Review's journals. 
\end{acknowledgments}



\clearpage

\appendix

\section{Identification and disambiguation of authors}
\label{sec:netw_id}

\begin{figure}[!tb]
\vskip .7cm
\includegraphics[width=0.45\textwidth]{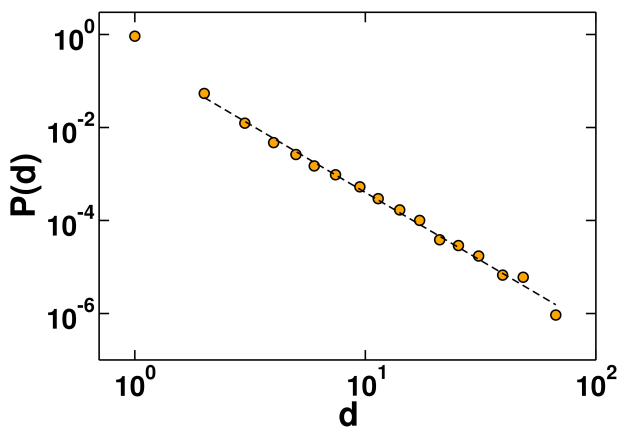}
\caption{(Color online) We consider only the IDs of authors with full version of their first names. Then, we count the 
number of times $d$ the same ID is obtained from authors with different first names 
(plus middle names, if present). The probability $P\left(d\right)$ (plotted as yellow circles) of finding an ID with ``degeneracy'' 
in the first name equal to $d$ has a power law decay as $d$ increases (the dashed line has exponent equal approximately to $-3$).}
\label{fig4_new}
\end{figure}

The list of references enables the construction of an error-free network of citation between articles. 
However, in this paper we are not interested in the analysis of
paper citation networks (PCNs), but on one of their particular projections: the Weighted Author Citation Network (WACN). We present a 
detailed description on the way in which we construct the WACN in 
section~\ref{sec:netw}. Here we would like to focus about possible 
sources of error, caused by the format of the PR dataset itself, 
associated with the projection of a network of citation between papers 
into the correspondent WACN.
\\
Whether authors can be well identified or not is still an open problem. Every author in the database has 
always a first and a last name. Many of them also have additional names, generically indicated as 
middle names. First (and middle) names may appear in their full version or they can only be represented 
by  the first letter. Writing first (and middle) names in their complete version is typically more common 
in recent papers and in papers with short lists of authors. On a total of $1\,916\,812$ repetitions for the 
authors (this means the sum of all authors, not only different authors, over all the papers) the first names 
appear $1\,564\,251$ times with just their first letter and the remaining $352\,561$ times in their full version.  
The simplest (and actually implemented) way to identify and distinguish authors is to assign to each author 
an identifier (ID) in accordance with the following rule


\begin{equation}
\left.
\begin{array}{l}
\textrm{\tiny LAST-NAME , F.  M.}
\\
\textrm{\tiny LAST-NAME , FIRST-NAME  MIDDLE-NAME}
\end{array}
\right\}
\Rightarrow
\textrm{\tiny LAST-NAME , FM} \;\;\; .
\label{eq:identifier}
\end{equation}

This means for example that according to rule~\ref{eq:identifier} ``Einstein, Abert'' has ID equal to ``Einstein, A'' 
while the ID of ``Bethe, Hans Albrecht'' is  ``Bethe, HA''. Essentially, the last name is taken in its full version, 
while for the first and the middle names we 
consider only the first letters. Proceeding in this way we are able to
distinguish  $216\,623$ ``different'' authors. 
\\
This approach is however biased by two main sources of error. First, there is a problem of identification for the 
authors. Unfortunately, scientists do not always sign their papers using the same name and this has as a consequence 
the impossibility to automatically relate different names to the same physical person. This fact may happen for 
several reasons: different order between first and last name; possible presence or absence of middle names; 
change of last names (this happens especially to ladies after their wedding).
\\
The second problem is basically the reverse of the formerly described source of error: the obvious impossibility 
to distinguish authors having same initials and the same last name by using only this information. We did not try 
to perform any kind of more elaborated analysis since this is still an open problem in bibliometrics and mainly 
because this was beyond the purposes of our paper. Furthermore, a simple analysis revealed that the number of 
``pathological'' cases is expected to  be small enough  to be considered irrelevant for the results reported in the paper. 
\\
In order to evaluate the relevance of the error introduced by the impossibility to disambiguate IDs, we consider only 
papers of our database signed by authors using the full version of their first and last names (and eventually their 
middle names). Unfortunately, this happens  only in recent papers (from $1980$ on) and only when the list of authors 
is sufficiently short (less than four, in general): this means that is very unlikely to happen. As already mentioned, 
the total number of ``signatures'' (i.e., the total number of non-distinct authors who have signed all papers in our 
database) is $1\,916\,812$, while the number of times in which an author has signed with her/his ``full signature'' 
is  only $352\,561$. Based on this subset, we perform the reduction described in rule~(\ref{eq:identifier}). We then 
calculate the probability $P\left(d\right)$ by simply counting the ratio between the total number of IDs shared by $d$ 
different scientists and the total number of IDs. The resulting distribution is plotted in Figure~\ref{fig4_new}: 
in the $92\%$ of the cases an ID corresponds to a single author; the rest of the distribution has a power law decay 
(i.e., $P\left(d\right) \sim d^{-\delta}$) as $d$ increases (the exponent $\delta \simeq 3$).

\begin{figure*}[!htb]
\vskip .7cm
\includegraphics[width=0.3\textwidth]{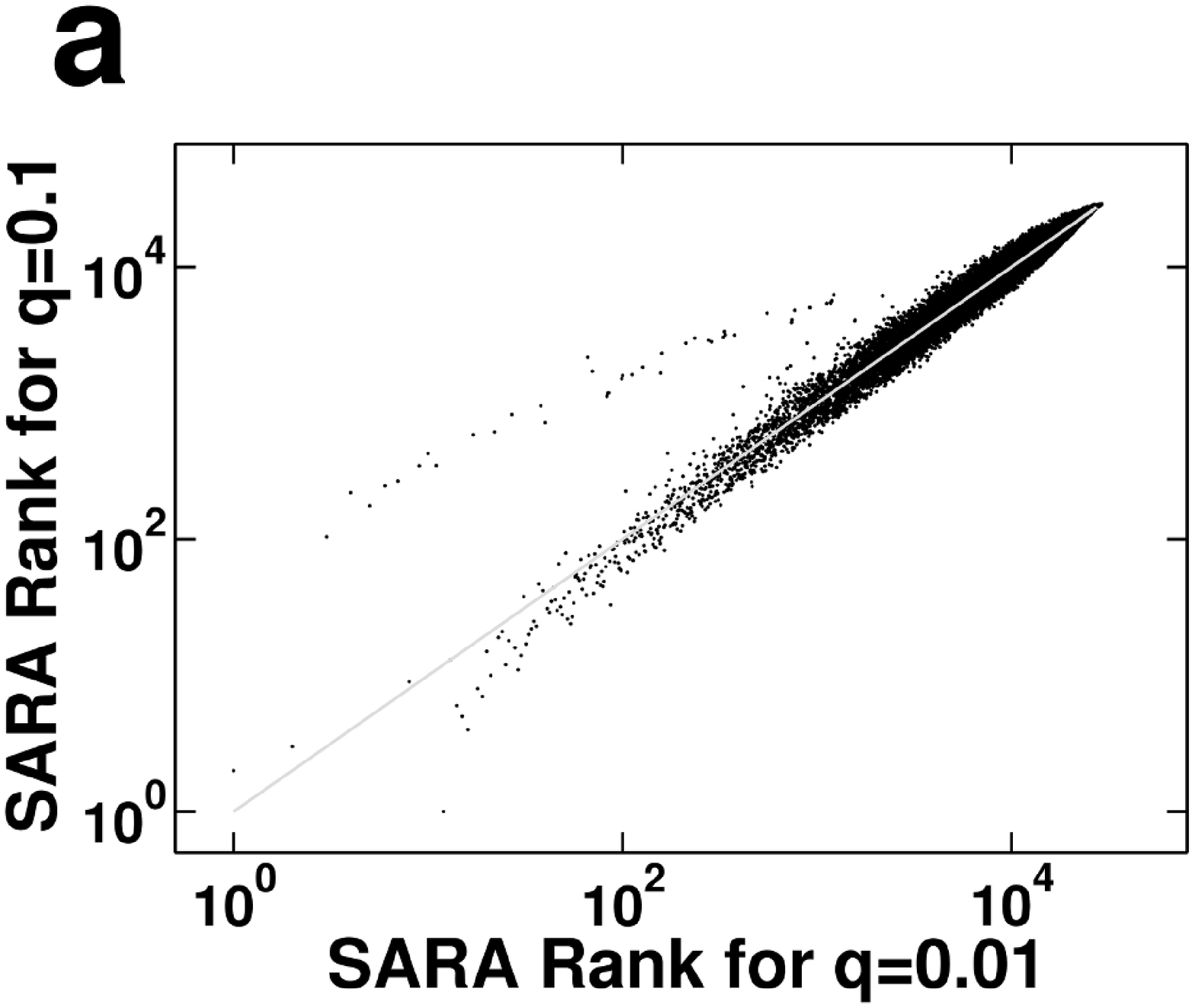}
\quad
\includegraphics[width=0.3\textwidth]{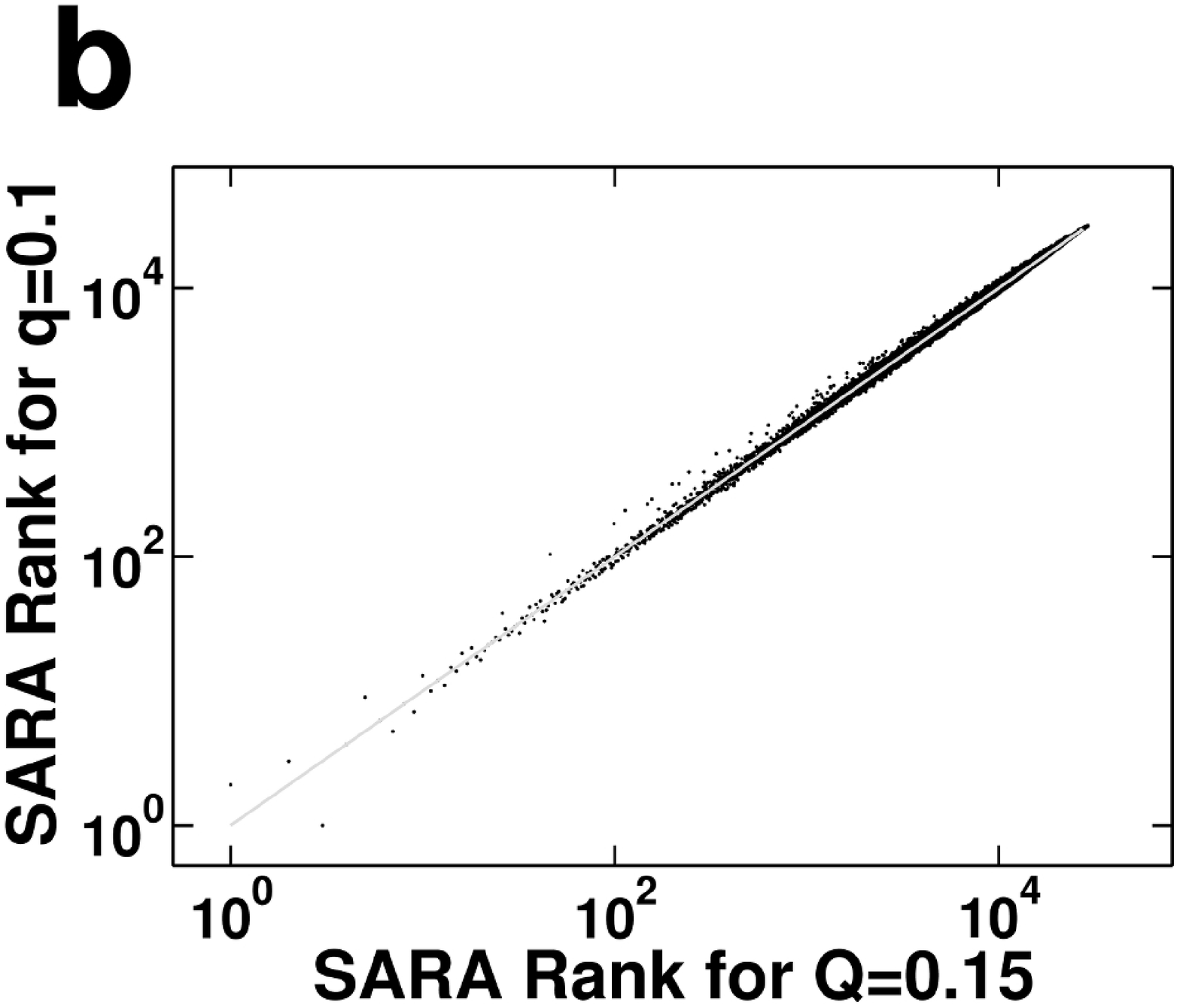}
\quad
\includegraphics[width=0.3\textwidth]{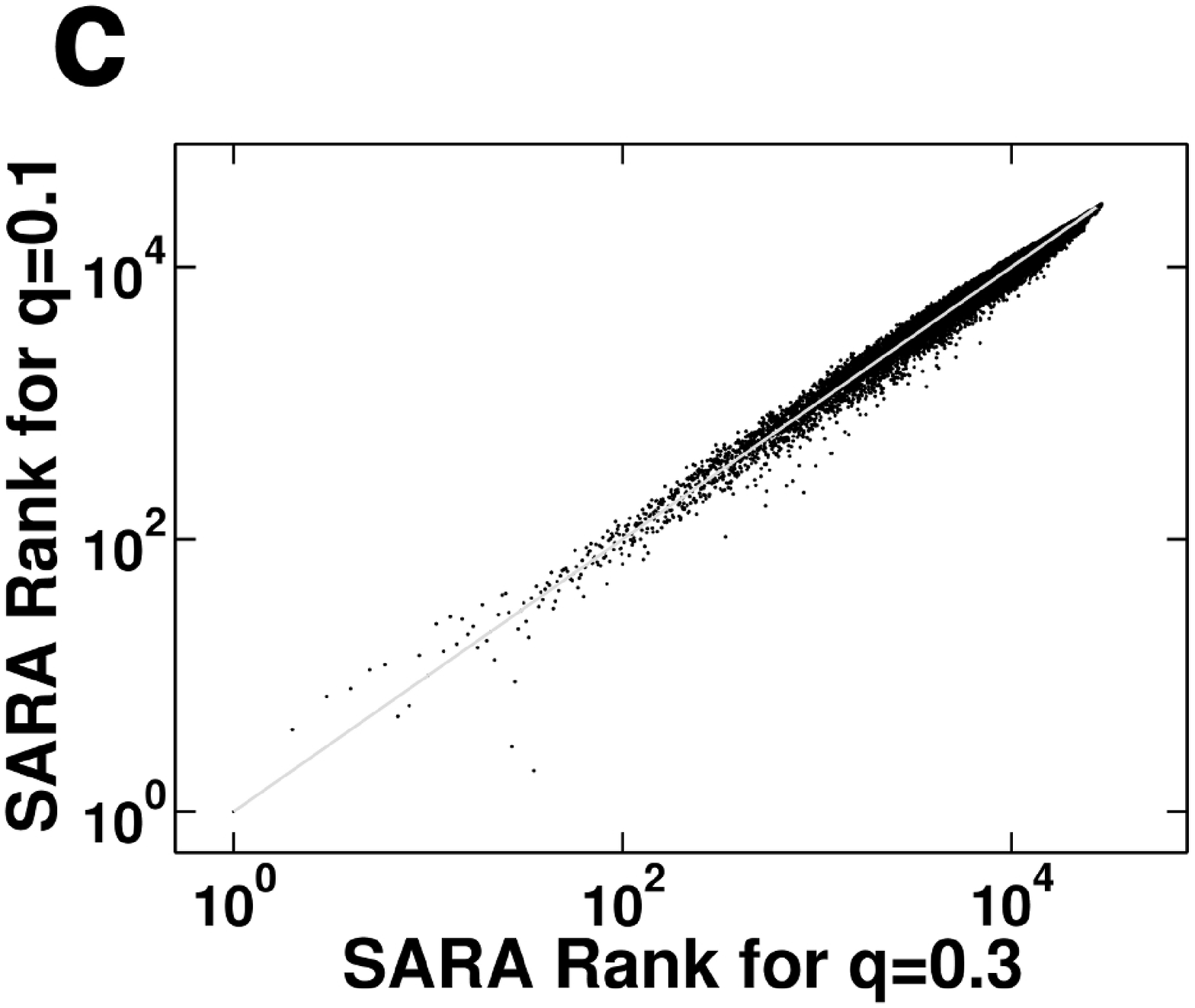}
\vskip .1cm
\caption{The rankings calculated with SARA for $q=0.1$ are plotted as function of the rankings obtained with the same algorithm but for different values of $q$: (a) $q=0.01$, (b) $q=0.15$ and (c) $q=0.3$. All plots have been generated from the WACN based on all papers published between $1893$ and $1966$ (the same dataset as the one used in Figures~\ref{fig6_new}a and~\ref{fig6_new}c of the main text).}
\label{fig1_app}
\end{figure*}

\section{Science Author Rank Algorithm: dependence on the damping factor}
\label{appendix}

Science Author Rank Algorithm (SARA) depends on the so-called damping factor $q$ [see Eq.~\ref{eq:pg}]. $q$ is a real number 
in the interval $[0,1]$ and the results calculated with SARA for different values of $q$ may differ. As a practical example, 
we report in Figure~\ref{fig1_app} some scatter plots between SARA rankings calculated for different values of $q$. 
As expected, SARA rankings calculated for different $q$ are linearly correlated and the correlation strength decreases 
as the difference between the $q$s increases.

\begin{figure}[!htb]
\vskip .7cm
\includegraphics[width=0.45\textwidth]{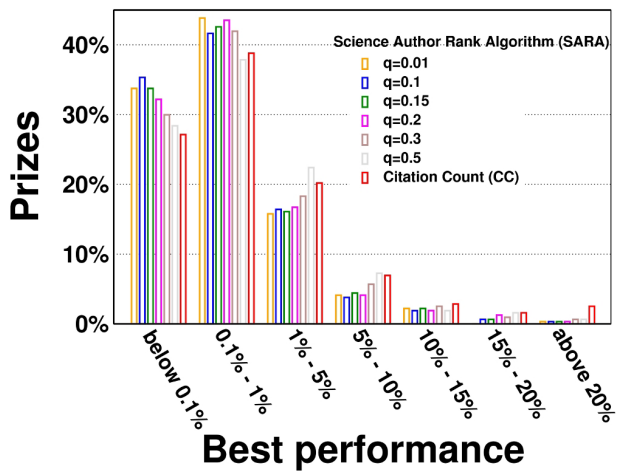}
\caption{(Color online) Percentage of prizes earned by physicists who have reached a given rank position as their best 
performance. Generally, the SARA is more predictive than the simple CC criterion since top scientists in 
SARA ranking have higher chances to earn a prize than top authors in the analogous ranking based on CC.}
\label{fig2_app}
\end{figure}

\

The decision to set $q=0.1$ is based on a special analysis which is graphically reported in Figure~\ref{fig2_app}. 
For each scientist, who earned one of the major prizes in Physics, we computed her/his best performance during 
her/his scientific history. We then plotted the ratio of prizes assigned to scientists with the best performance 
falling in a given interval (note that the intervals' division is totally arbitrary, but the results do not strictly 
depend on this choice). According to any reasonable measure of scientific impact, the probability that a scientist 
earns an important prize should be related to her/his scientific relevance. In the case of SARA ranking, we generally 
observed that the majority of prizes is assigned to scientists who have reached a top position in the ranking. 
This allows us to justify the use of such measure for the scientific impact of authors. Moreover, as already 
stated and shown (see Figure~\ref{fig7_new}), SARA is more effective than other well known criteria like 
Citation Count (CC) or Balanced Citation Count (BCC) if one wants to predict future winners of prizes. 
Anyway, also in the case of SARA, the predictivity of the algorithm may quantitatively change as function 
of $q$. Looking at Figure~\ref{fig2_app}, we see for instance that, in the top intervals, the highest 
ratios are reached for values of $q \simeq 0.1$, while values of $q<0.1$ or $q>0.1$ give lower ratios 
in these first two bins. As a consequence, we can say that $q=0.1$ is the optimal value for SARA since 
it is the value which maximizes the predictivity of our algorithm.


\begin{thebibliography}{50}

\bibitem{egghe}
L.~Egghe \& R.~Rousseau,
{\it Introduction to Informetrics:
quantitative methods in library, documentation
and information science}, (Elsevier, Amsterdam, 1990).

\bibitem{garfield}
E.~Garfield, {\it Citation Indexing. Its Theory and
Applications in Science, Technology, and Humanities}, (Wiley, New York,
1979).


\bibitem{adler08}
R.~Adler, J.~Ewing \& P.~Taylor, {\it IMU Report: Citation Statistics},
{\it {\scriptsize http://www.mathunion.org/Publications/Report/CitationStatistics}} (2008).


\bibitem{hirsch05}
J.~E.~Hirsch, 
Proc. Natl. Acad. Sci. USA {\bf 102}, 16569-16572 (2005).



\bibitem{Newman:2001} M.~E.~J.~Newman, 
Proc. Natl. Acad. Sci. USA {\bf 98}, 404-409 (2001).

\bibitem{Newman:2001a} M.~E.~J.~Newman,  
Phys. Rev. E {\bf 64}, 016131 (2001).

\bibitem{Newman:2001b} M.~E.~J.~Newman,  
Phys. Rev. E {\bf 64}, 016132 (2001).

\bibitem{vicsek:02}
A.L.~Barab\'asi, H.~Jeong, Z.~Neda, E.~Ravasz, A.~Schubert \& T.~Vicsek,
Physica A {\bf 311}, 590-614 (2002).

\bibitem{Redner98}
S.~Redner, 
Eur. Phys. J. B {\bf 4}, 131-134 (1998).


\bibitem{chen07}
P.~Chen, H.~Xie, S.~Maslov,\& S.~Redner, 
Journal of Informetrics {\bf 1}, 8-15 (2007).



\bibitem{brin98}
S.~Brin \& L.~Page,
Computer Networks and ISDN Systems {\bf 30}, 107-117 (1998).


\bibitem{kleinberg99}
J.~Kleinberg, 
Journal of the ACM {\bf 46}, 604 (1999).



\bibitem{donato}
C.~Castillo, D.~Donato \& A.~Gionis, 
Lecture Notes in Computer Science, (Springer-Verlag, Berlin, 2007). 

\bibitem{sidiropoulos}
A.~Sidiropoulos \& Y.~Manolopoulos, 
Journal for Systems \& Software {\bf 79}, 1679-1700 (2006).


\bibitem{walker07}
D.~Walker, H.~Xie, K.~K.~Yan \& S.~Maslov,
J. Stat. Mech. P0610 (2007).

\bibitem{redner05}
S.~Redner, 
Phys. Today {\bf 58}, 49-54 (2005).


\bibitem{barrat04}
A.~Barrat, M.~Barth\'elemy, R.~Pastor-Satorras \& A.~Vespignani, 
Proc. Natl. Acad. Sci. USA {\bf 101},  3747-3752 (2004).


\bibitem{radicchi08}
F.~Radicchi, S.~Fortunato \& C.~Castellano, 
Proc. Natl. Acad. Sci. USA {\bf 105},   17268-17272 (2008).


\bibitem{fortunato08}
S.~Fortunato, M.~Boguna, A.~Flammini \& F.~Menczer, 
Proc. WAW 2006 LNCS {\bf 4936},  59-71 (2008).

\bibitem{garfield86}
E.~Garfield,  
Essays of an Information Scientist {\bf 4}, 182-187 (1986).


\bibitem{Enserink09}
M.~Enserink,
Science {\bf 323}, 1662-1664 (2009).

\end{thebibliography}
\end{document}